\documentclass{aa}
\usepackage{graphics}
\begin{document}

%

%
   \title{The ``red shelf" of the H$\beta$ line in the Seyfert 1 galaxies
RXS~J01177+3637 and HS~0328+05}
   \subtitle{}

   \author{P. V\'eron \inst{1}, A.C. Gon\c{c}alves \inst{2} and M.-P. 
V\'eron-Cetty \inst{1}
\thanks{ \small Based on observations obtained with the ESO NTT telescope, La 
Silla, Chile and the 1.93-m telescope at the Observatoire de Haute Provence 
(CNRS), France }}

\offprints{P. V\'eron}

\institute{
Observatoire de Haute Provence, CNRS, F-04870 Saint-Michel l'Observatoire,
    France\\ 
\email{veron@obs-hp.fr; mira@obs-hp.fr}
\and
European Southern Observatory (ESO), Karl Schwarzschild Strasse 2, D-85748 
Garching bei M\"unchen, Germany\\
\email{adarbon@eso.org}}

\date{Received 19 November 2001 / Accepted 9 January 2002}

\abstract{ A few Seyfert 1s have a H$\beta$ profile with a red wing usually 
called the ``red shelf". The most popular interpretation of this feature is that
it is due to broad redshifted lines of H$\beta$ and 
[O~III]$\lambda\lambda$4959,5007; we have observed two Seyfert 1s displaying a 
``red shelf" and showed that in these two objects the main contributor is most 
probably the He~I $\lambda\lambda$4922,5016 lines having the velocity and width 
of the broad H$\beta$ component.
 There is no evidence for the presence of a broad redshifted component of
H$\beta$ or [O~III] in any of these two objects.
      \keywords{galaxies: Seyfert--galaxies: individual RXS~J01177+3637,
HS~0328+05}}

\titlerunning{Red shelf in RXS~J01177+3637 and HS~0328+05}
\authorrunning{V\'eron et al.}
   \maketitle



\section{Introduction}

 A few Seyfert 1s have a very complex H$\beta$ profile with a strong red wing
extending underneath the [O~III]$\lambda\lambda$4959,5007 lines. The excess 
emission in the red wing is 
referred to as the ``shelf" feature or the ``red shelf" (Meyers \& Peterson 
\cite{meyers}). In most Seyfert 1s showing such a feature, it appears to be made
of two components: a broad red wing to H$\beta$, and a broad wing on the long 
wavelength side of the [O~III]$\lambda$5007 line (van Groningen \& de Bruyn 
\cite{groningen}).

 Several interpretations have been proposed to explain the ``red shelf": it 
could be due to H$\beta$ or to the presence of other broad emission lines such 
as [O~III], Si~II $\lambda$5056, He~I $\lambda$5016 (Meyers \& Peterson 
\cite{meyers}; van Groningen \& de Bruyn \cite{groningen}; Kollatschny et al.
\cite{kollatschny01}) or Fe~II (Korista \cite{korista92}).

 Meyers \& Peterson (\cite{meyers}), Crenshaw \& Peterson (\cite{crenshaw1986}) 
and Stirpe et al. (\cite{stirpe89}) have argued that broad [O III] lines are 
most probably the main contributor to the observed $\lambda$5007 red wing. Van 
Groningen \& de Bruyn (\cite{groningen}) have found the same red wing in 
[O~III]$\lambda$4363 in several objects, confirming that they are indeed due to
[O~III] emission.

 To investigate the nature of the ``red shelf", we have made spectroscopic 
observations of two Seyfert 1s: RXS~J01177+3637 and HS~0328+05.

 \section{Observations and data reduction}

 RXS~J01177+3637 was observed on January 15, 1999, with the CARELEC spectrograph
(Lema\^{\i}tre et al. \cite{lemaitre89}) attached to the Cassegrain focus of the
OHP 1.93-m telescope. We obtained one 20 min exposure with a dispersion of 130 
\AA\ mm$^{-1}$ in the range 4600 to 7900 \AA. The detector was a 
1024$\times$2048, 13.5$\times$13.5 $\mu$m pixel EEV CCD. Nine lines were 
extracted. The slit width was 2\farcs2 corresponding to 4.0 pixels on the 
detector; the resolution, as measured on the night sky lines, was $\sim$6 \AA\ 
FWHM. The spectrum was flux calibrated with the standard stars EG~166 and EG~247
(Oke \cite{oke}) and Feige~66 (Massey et al. \cite{massey}).

\begin{figure}[ht]
\resizebox{8.8cm}{!}{\includegraphics{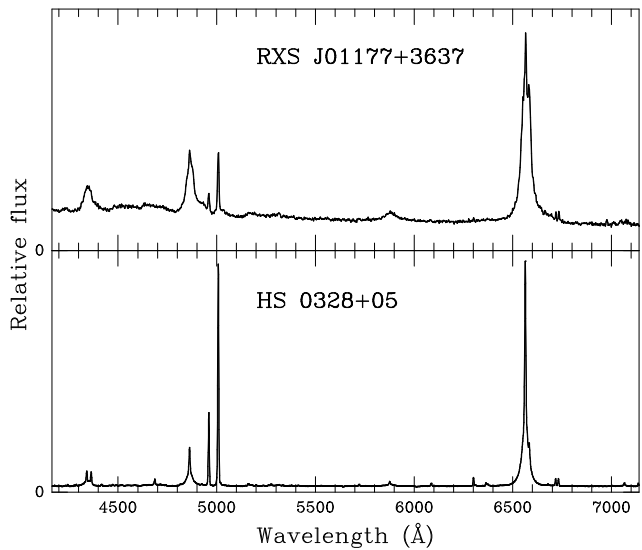}}
\caption{\label{deuxsp}
Deredshifted spectra of RXS~J011177+3637 and HS~0328+05.
}
\end{figure}

 In the case of HS~0328+05, the observations were carried out with the EMMI 
spectrograph attached to one of the Nasmyth foci of the ESO 3.58-m NTT telescope
at La Silla. The detector was a 2048$\times$2047, 24$\times$24 $\mu$m pixel 
Tektronix CCD.
 Two spectra with a high signal-to-noise ratio ($>$50 in the continuum) were 
obtained, one in the red on November 24, 2000 (grism\#6,   
6000-8300 \AA), the other in the blue on November 23 (grism\#5,
4000-6600 \AA). The exposure times were 30 and 45 min 
respectively.
 The slit width was 1\arcsec, corresponding to 3.7 pixels on the detector; the 
resolution, as measured on the night sky 
lines, was $\sim$ 4.5 \AA\ FWHM. The nucleus was centered on the slit which was
aligned with the parallactic angle. Seven lines were extracted. The spectra were
flux calibrated with the standard stars LTT~2415, LTT~3218 and LTT~9491 taken 
from Hamuy et al. (1992, 1994).

 The spectra of the two objects are shown in Fig. \ref{deuxsp}. All the 
following analysis was done using our own software (V\'eron et al. \cite{veron80}).

\section{Analysis}

\subsection{RXS~J01177+3637}
 
 RXS~J01177+3637 is a Seyfert 1 at z=0.106 (Wei et al. \cite{wei99}). Our
spectrum clearly shows the presence of the ``red shelf" (Fig. \ref{deuxsp}). \\

 In a first step the broad H$\alpha$ line was fitted with two Gaussians; the 
first (G1) has a FWHM of 2\,150 km s$^{-1}$ and a velocity of 130 km s$^{-1}$ 
(with respect to the narrow emission lines), while the second (G2) has a peak 
intensity of 24\% of that of the first, a FWHM of 6\,100 km s$^{-1}$ and a 
velocity of 560 km s$^{-1}$.

\begin{figure}[ht]
\resizebox{8.8cm}{!}{\includegraphics{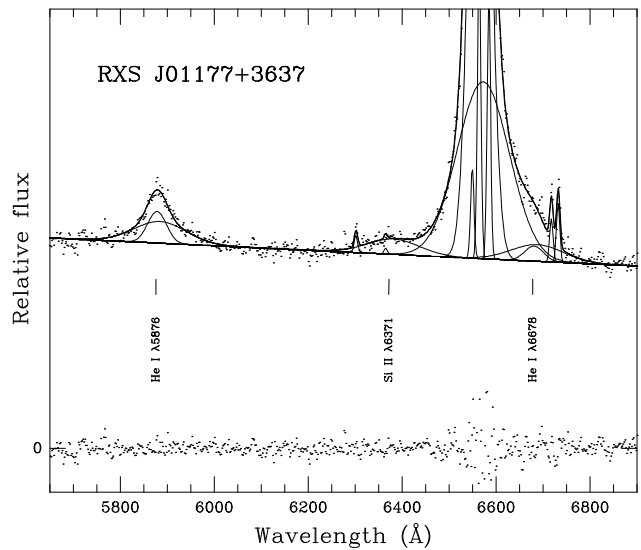}}
\caption{\label{rxsj_ha}
Fit of the red part of the spectrum of RXS~J01177+3637. The figure shows the 
data, the fit, the individual components and the residuals.
}
\end{figure}

 The red part of the spectrum also shows broad He~I lines: $\lambda$5876, 
$\lambda$6678, $\lambda$7065 and a line near $\lambda$6370. The three He~I lines 
have the width of the narrower H$\alpha$ 
component G1 ($\lambda$5876/H$\alpha$=4.1\%, $\lambda$6678/H$\alpha$=2.0\%; 
$\lambda$7065/H$\alpha$=2.6\%; here the H$\alpha$ flux is that of the G1
component. The ionization potentials of Si~I, Si~II and Si~III are nearly identical
to the ionization potentials of Fe~I, Fe~II and Fe~III and therefore the Fe$^{+}$ 
and Si$^{+}$ zones should be virtually the same; consequently Si~II lines are 
expected in objects with strong Fe~II emission (Phillips \cite{phillips78}); in
several Seyfert 1s, an emission feature near 5050 \AA\ has been attributed to
Si~II $\lambda$5056 (Crenshaw\& Peterson \cite{crenshaw1986}); we therefore feel 
confident that the line observed at $\lambda$6370 is Si~II $\lambda$6371; this
line has the width of the broader H$\alpha$ component G2 
($\lambda$6371/H$\alpha$=8.6\%), while the He~I $\lambda$5876 and $\lambda$6678 
lines also have a broader component ($\lambda$5876/H$\alpha$=12\% and 
$\lambda$6678/H$\alpha$=9.5\%). Fig. \ref{rxsj_ha} shows the result of the fit 
of the red part of the spectrum. \\

\begin{figure}[ht]
\resizebox{8.8cm}{!}{\includegraphics{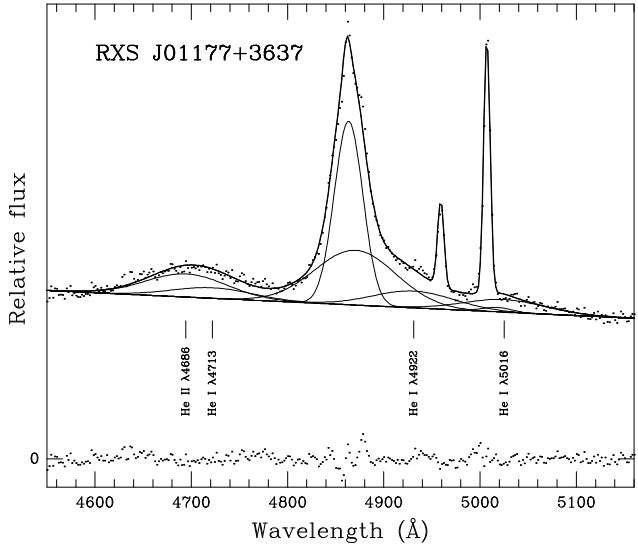}}
\caption{\label{rxsj_hb}
Fit of the blue part of the spectrum of RXS~J01177+3637. The figure shows the 
data (after subtraction of the Fe~II emission), the fit, the individual broad 
components (H$\beta$, He~I $\lambda$4713, $\lambda$4922 and $\lambda$5016 and 
He~II $\lambda$4686) and the residuals. }
\end{figure}

  In a second step we removed the Fe~II multiplets in the blue part of the 
spectrum following the method described by Boroson \& Green (\cite{borogr}), 
using a Fe~II template obtained by taking a high signal-to-noise spectrum of 
I~Zw~1 (V\'eron-Cetty et al. \cite{veron01}), an NLS1 showing strong narrow 
Fe~II emission. The resulting spectrum still shows strong broad red wings to 
H$\beta$ and $\lambda$5007 (Fig. \ref{rxsj_hb}).

 We fitted the broad H$\beta$ line with two Gaussian profiles having the 
velocities and widths of the Gaussian components (G1 and G2) of the broad 
H$\alpha$ line. The broad red wings to H$\beta$ and $\lambda$5007 are well 
fitted with broad (FWHM=6\,100 km s$^{-1}$) Gaussian profiles (G2) at the 
wavelengths of He~I $\lambda$4922 and $\lambda$5016 
($\lambda$4922/H$\beta$=30\%; $\lambda$5016/H$\beta$=22\%); the $\lambda$5016 
line may also have a narrower (2\,150 km s$^{-1}$ FWHM) component (G1) 
with $\lambda$5016/H$\beta$=2.4\%. The He~II 
$\lambda$4686 and He~I $\lambda$4713 lines are fitted each with a Gaussian 
profile with FWHM=6\,100 km s$^{-1}$ (G2); the He~II line is very strong 
($\lambda$4686/H$\beta$=42\%); $\lambda$4713/H$\beta$=19\% 
(Fig.~\ref{rxsj_hb}). \\

 Although higher quality spectra would be necessary to conclude unambiguously,
it seems that the red wings to H$\beta$ and $\lambda$5007 may be fully accounted
for by the presence of the broad He~I $\lambda$4922 and $\lambda$5016 lines
respectively.

\subsection{HS~0328+05}

 HS~0328+05 is a Seyfert 1 at z=0.043 (Perlman et al. \cite{perlman96}; Engels
et al. \cite{engels98}). The Balmer lines have been fitted with a broad 
Lorentzian profile with a FWHM of $\sim$1\,500 km s$^{-1}$ leading to the 
classification of this object as a NLS1; the Fe~II emission was not detected, 
with the ratio of the Fe~II emission measured between 4450 and 4684 \AA\ and the
total H$\beta$ flux R$_{4570}<$0.6 (V\'eron-Cetty et al. 
\cite{veron01}); our new high quality spectrum shows that H$\beta$ has a weak, 
but definitively present, ``red shelf"; in addition, the Fe~II emission is 
detected with R$_{4570}$=0.43.

\begin{figure}[ht]
\resizebox{8.8cm}{!}{\includegraphics{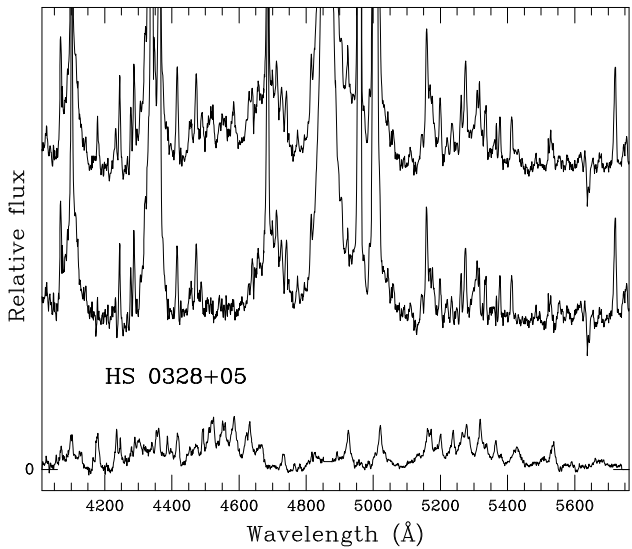}}
\caption{\label{hsfe}
The upper curve is the spectrum of HS~0328+05; the lower curve is the suitably 
scaled template Fe~II spectrum; the middle curve is the difference (moved 
downward by an arbitrary amount for clarity). The quality of the iron removal is
well seen in the spectral range 4450 to 4650 \AA. }
\end{figure}

\subsubsection{Narrow emission lines in the range 5080-5430~\AA}

 In a first step we removed the broad Fe~II multiplets as described above. In 
so doing we made the implicit assumption that the Fe~II spectra of HS~0328+05 
and I~Zw~1 are identical; however according to Joly (\cite{joly88}) the relative
intensities of the various multiplets are not the same in all objects, being a 
function of the temperature. Is then our assumption justified?

\begin{figure}[ht]
\resizebox{8.8cm}{!}{\includegraphics{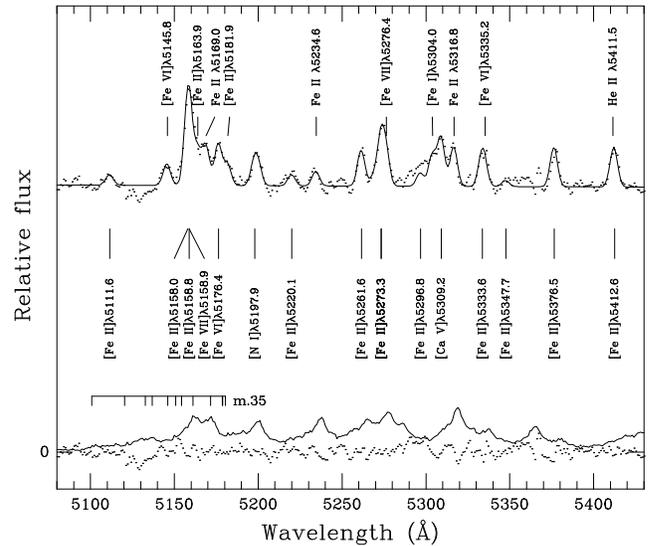}}
\caption{\label{hs_feiilines}
Spectrum of HS~0328+05 (after subtraction of the broad permitted Fe~II lines) 
in the range 5080 to 5430 \AA. All identified emission features were included in
the fit assuming that they all have the same velocity and width (namely 4.9 
\AA\ FWHM). Thirteen forbidden and three permitted Fe~II emission lines are 
clearly detected. 
The bottom dotted line shows the residuals, while the solid line is the Fe~II 
template subtracted from the original spectrum. The positions of the twelve 
lines in Fe~II multiplet 35 are indicated. 
}
\end{figure}

 Inspection of the I~Zw~1 Fe~II spectrum (Fig.~\ref{hsfe}) shows that, in this 
object, the intensities of multiplets 25 (4846-5000 \AA), 35 
(5100-5180 \AA) and 36 ($\lambda\lambda$4893,4993,5037) are 
quite small compared to that of m. 42 ($\lambda\lambda$4924,5018,5169), thus
suggesting a
temperature T$_{\rm e}>$ 8\,000 K; if the temperature in HS~0328+05 were smaller
(T$_{\rm e}<$ 7\,000 K) the relative intensity of the first three multiplets
would be much larger with respect to m. 42 (Joly \cite{joly88}). Inspection of 
the residuals in Fig. \ref{hs_feiilines} shows that the Fe~II m. 35 has been 
completely cancelled out from the spectrum of HS~0328+05 by subtracting the 
template. It seems therefore that our procedure has indeed removed all the Fe~II
emission and that our initial assumption is justified. \\

\begin{table}[!ht]
\caption{\label {FeII} [Fe~II] lines observed in HS~0328+05. Col. 2 gives the 
multiplet numbers, col. 3 the radiative transition probabilities A (from Quinet 
et al. 1996), col. 4 the line intensities measured in the Orion nebula relative 
to He~I $\lambda$6678, multiplied by 100 (Verner et al. 2000), col. 5 the 
relative intensities measured in HS~0328+05, col. 6 the log of the ratio of 
the intensities observed in HS~0328+05 and Orion.}
\begin{center}
\begin{tabular}{lrlllr}
\hline
$\lambda_{\rm lab}$ (\AA) & & A & I(Orion) & I(HS) \\
\hline
 4114.5 & 23F & 0.103 & 0.25 & 0.30 &   0.08 \\
 4177.2 & 21F & 0.194 & 0.26 & 0.23 & --0.05 \\
 4244.0 & 21F & 1.12  & 1.18 & 1.23 &   0.06 \\
 4276.8 & 21F & 0.819 & 0.92 & 0.77 & --0.08 \\
 4287.4 &  7F & 1.65  & 2.32 & 1.58 & --0.17 \\
 4319.6 & 21F & 0.658 &      & 0.25 &        \\
 4352.8 & 21F & 0.380 & 0.51 & 0.45 & --0.05 \\
 4358.4 & 21F & 0.875 &      & 1.98 &        \\
 4359.3 &  7F & 1.22  & 1.60 & 1.36 & --0.07 \\
 4372.4 & 21F & 0.340 &      & 0.20 &        \\
 4413.8 &  7F & 0.858 & 1.16 & 0.82 & --0.16 \\
 4416.3 &  6F & 0.454 & 1.41 & 0.73 & --0.28 \\
 4452.1 &  7F & 0.548 & 0.90 & 0.53 & --0.23 \\
 4457.9 &  6F & 0.279 & 0.49 & 0.56 &   0.06 \\
 4474.9 &  7F & 0.267 & 0.35 & 0.26 & --0.13 \\
 4488.7 &  6F & 0.150 &      & 0.55 &        \\
 4639.7 &  4F & 0.499 &      & 0.58 &        \\
 4664.4 &  4F & 0.156 &      & 0.18 &        \\
 4728.1 &  4F & 0.478 & 0.16 & 0.56 &   0.54 \\
 4774.7 & 20F & 0.163 & 0.17 & 0.34 &   0.30 \\
 4814.5 & 20F & 0.521 & 1.20 & 1.01 & --0.07 \\
 4874.5 & 20F & 0.223 &      & 1.16 &        \\
 4889.6 &  4F & 0.347 & 0.67 & 1.25 &   0.27 \\
 4905.3 & 20F & 0.285 & 0.23 & 0.53 &   0.36 \\
 4947.4 & 20F & 0.075 & 0.49 & 0.53 &   0.03 \\
 4950.7 & 20F & 0.225 & 0.35 & 0.24 & --0.16 \\
 4973.4 & 20F & 0.183 & 0.36 & 0.17 & --0.33 \\
 5020.2 & 20F & 0.236 &      & 0.25 &        \\ 
 5043.5 & 20F & 0.094 &      & 0.49 &        \\
 5111.6 & 19F & 0.131 & 0.38 & 0.23 & --0.22 \\
 5158.0 & 18F & 0.440 &      & 0.62 &        \\
 5158.8 & 19F & 0.605 & 1.82 & 1.25 & --0.16 \\
 5163.9 & 35F & 0.309 &      & 0.67 &        \\
 5181.9 & 18F & 0.500 &      & 0.51 &        \\
 5220.1 & 19F & 0.144 & 0.15 & 0.27 &   0.26 \\
 5261.6 & 19F & 0.429 & 1.36 & 0.68 & --0.30 \\
 5268.9 & 18F & 0.288 & 0.09 & 0.21 &   0.37 \\
 5273.3 & 18F & 0.550 & 0.67 & 1.16 &   0.24 \\
 5296.8 & 19F & 0.118 & 0.08 & 0.54 &   0.83 \\
 5333.6 & 19F & 0.351 & 0.35 & 0.62 &   0.25 \\
 5347.7 & 18F & 0.087 &      & 0.12 &        \\
 5376.4 & 19F & 0.348 & 0.46 & 0.78 &   0.23 \\
 5412.6 & 17F & 0.283 &      & 0.63 &        \\
 5527.3 & 17F & 0.273 &      & 0.46 &        \\ 
 5747.0 & 34F & 0.373 & 0.08 & 0.49 &   0.79 \\
 7155.2 & 14F & 0.153 & 1.50 & 0.99 & --0.18 \\
 7172.0 & 14F & 0.588 & 0.35 & 0.62 &   0.25 \\
 7388.2 & 14F & 0.045 & 0.24 & 0.24 &   0.00 \\
 7452.5 & 14F & 0.049 & 0.45 & 0.44 & --0.01 \\
 
\hline
\end{tabular}
\end{center}
\end{table}
\normalsize

 The resulting spectrum is rich in narrow emission lines. We analysed a region 
free of broad emission features: 5080-5430 \AA. 
Fig.~\ref{hs_feiilines} shows all the lines 
identified in this range including thirteen [Fe~II] lines, three Fe~II lines 
($\lambda$5169 (42), $\lambda$5234 (49) and $\lambda$5317 (48)), three [Fe~VI]
(2F) lines at $\lambda$5145.8, $\lambda$5176.4 and $\lambda$5335.2, and two 
[Fe~VII] (2F) lines at $\lambda$5158.9 and $\lambda$5276.4.

 The observed [Fe~VI] lines have large predicted intensities (Garstang et al. 
\cite{garstang78}).

 The two [Fe~VII] lines $\lambda\lambda$5721,6087 (1F) are relatively strong
with a ratio $\lambda$6087/$\lambda$5721=1.53, near the theoretical value of
1.60 (Nussbaumer \& Storey \cite{nussbaumer}). According to Keenan \& Norrington
(\cite{keenan87}), for densities lower than $\sim$ 10$^{6}$ cm$^{-3}$, the
intensity ratio $\lambda$5159/$\lambda$6087 is in the range 0.20-0.35. In the 
Seyfert 1 III~Zw~77, Osterbrock (\cite{osterbrock81}) has observed 
$\lambda$5159/$\lambda$6087=0.20. The two [Fe~VII] lines $\lambda$5159 and 
$\lambda$5276 are blended with [Fe~II] $\lambda$5159 (19F) and $\lambda$5273
(18F) respectively; we could however determine the intensity of
[Fe~VII]$\lambda$5276; assuming that the intensities of $\lambda$5276 and
$\lambda$5159 are equal (their radiative transition probabilities are about 
equal, Keenan \& Norrington \cite{keenan87}), we got 
$\lambda$5159/$\lambda$6087=0.2, confirming the identification of these two 
blended lines. \\

 Having shown that the spectrum is dominated by [Fe~II] lines, we looked for 
lines from this ion in the whole spectrum; the lines found are listed in Table 
\ref{FeII} with their radiative transition probabilities and their relative 
intensities both in Orion when available (Verner et al. \cite{verner00}) and 
HS~0328+05. These intensities are in good agreement (within a factor of 2),
except for the weakest lines (I$_{\rm Orion}<$0.25) for which our intensities 
are larger. A few of these lines were previously observed in the spectrum of 
NGC~4151 (Boksenberg et al. \cite{boksenberg}). Several of them were not 
observed neither by Verner et al. (Orion) nor by Hamann (\cite{hamann94}) 
(KK~Oph); however their radiative transition probabilities computed by Quinet et
al. (\cite{quinet}) are relatively large. \\

 \subsubsection{Fitting the spectrum around H$\alpha$ and H$\beta$}

\begin{figure}[ht]
\resizebox{8.8cm}{!}{\includegraphics{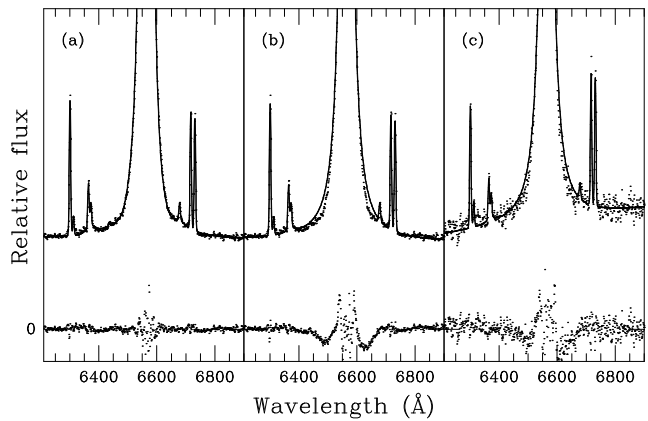}}
\caption{\label{hs_ha}
 Fit of the spectrum of HS~0328+05 in the range 6210 to 6900 \AA. The 
broad H$\alpha$ component has been fitted with three Gaussian profiles (a),
with a single Lorentzian (b). (c) is the lower signal-to-noise ratio spectrum 
obtained with the OHP 1.93-m telescope fitted with a single Lorentzian. The 
high signal-to-noise reached with the NTT spectrum shows that the single 
Lorentzian fit is no longer acceptable. }

\end{figure}

 We fitted the red spectrum (6250-6740 \AA) with Gaussian 
components. The broad H$\alpha$ is well fitted by three Gaussians H1, H2 and 
H3; their velocities are 51, --98 and 29 km s$^{-1}$ respectively and their FWHM
1\,730, 3\,400 and 11\,600 km s$^{-1}$. The relative peak intensity of 
these three components is 1.00, 0.40 and 0.04.  
  The He~I $\lambda$6678 line has a weak broad (1\,730 km s$^{-1}$ FWHM) 
component (H1), with $\lambda$6678/H$\alpha$=0.9\%; broad lines (H1) of He~I 
$\lambda$5876 and $\lambda$7065 are also present ($\lambda$5876/H$\alpha$=2.1\%,
$\lambda$7065/H$\alpha$=2.4\%); $\lambda$5876 has also a broader (3\,400 km 
s$^{-1}$ FWHM) component (H2) with $\lambda$5876/H$\alpha$=5.5\%. A broad (H1) 
line is required at $\lambda$6371 that we identify with Si~II $\lambda$6371. \\

 We showed in a preceding paper (V\'eron-Cetty et al. \cite{veron01}) that the 
Balmer lines of NLS1s are better fitted with a single Lorentzian profile 
than with a single Gaussian profile. Our new high signal-to-noise spectrum shows
that, at least in the case of HS~0328+05, this is not satisfactory and that a 
fit involving several Gaussians is needed (see Fig.~\ref{hs_ha}). \\

 We analysed in detail the blue spectrum in the range 4780 to 5120 \AA\ 
which contains the H$\beta$ and [O~III] lines.

\begin{figure}[ht]
\resizebox{8.8cm}{!}{\includegraphics{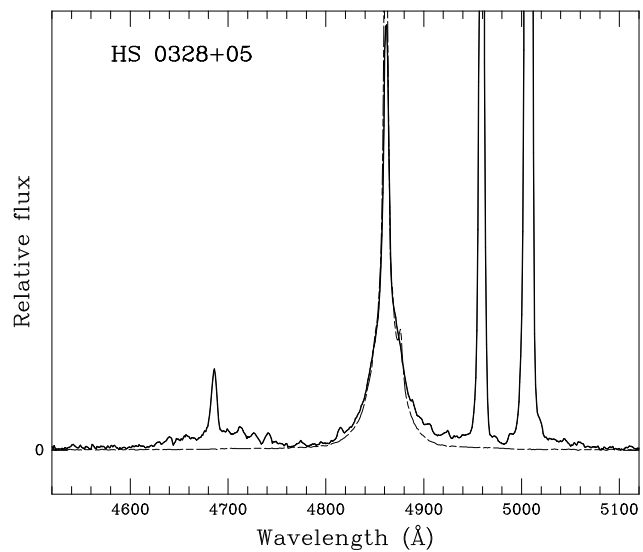}}
\caption{\label{hs_hahb}
The solid line is the blue spectrum of HS~0328+05 after subtraction of the Fe~II
emission; the dashed line is the H$\alpha$ profile moved to the wavelength of 
H$\beta$. The excess of emission redward of H$\beta$ is obvious.
}
\end{figure}

  Fig.~\ref{hs_hahb} shows the H$\alpha$ profile superposed on the H$\beta$ 
profile. The presence of an excess of emission in the red wings of H$\beta$ and
[O~III]$\lambda$5007 is clearly visible. These broad emission features, centered
at $\sim\lambda$4890 and $\sim\lambda$5050, are separated by a significant deep 
at $\sim\lambda$4980 suggesting that they are two distinct features.

 We have divided the H$\alpha$ profile by the H$\beta$ profile; the 
H$\alpha$/H$\beta$ ratio is $\sim$4 at the line center but drops quickly to 
$\sim$1 in the region of the red wing (4880-4940 \AA), an 
unlikely value for the Balmer decrement. It seems therefore that the feature 
seen as a red wing to H$\beta$ cannot be due to H$\beta$ emission. \\

 For the blue spectrum, we made a fit with Gaussian components including, for 
H$\beta$, two broad components having the velocity and FWHM of the two main 
broad components of H$\alpha$ (H1 and H2); their intensity was let free; we 
found the Balmer decrement for these two components to be equal to 5.5 and 3.7 
respectively. If the third, very broad, component (H3) had a Balmer decrement 
larger than three, it would not be detectable at H$\beta$.

 We found necessary to include two broad Gaussian components to each of the He~I
lines $\lambda$4922 and $\lambda$5016, having the velocity and width of the
narrowest broad Balmer profiles H1 and H2 respectively; their intensity with 
respect to the H$\beta$ components is 8.1\% and 3.8\% respectively for H1 and
2.0\% and 18.7\% for H2. \\

 We tried to identify all narrow lines in the range 4780 to 5120 \AA.
In addition to the [Fe~II] lines listed in Table\,\ref{FeII}, we have identified
two narrow permitted Fe~II (42) lines: $\lambda$4924, $\lambda$5018, as well as 
[Fe~VII] $\lambda$4893, $\lambda$4942 and $\lambda$4989, three lines in 
multiplet 2F with relatively large radiative transition probalities, and 
[Fe~VI] $\lambda$4967 and $\lambda$4972 (2F), two lines with relatively strong 
predicted intensities (Garstang et al. \cite{garstang78}). 

 Several [Fe~IV] lines have been observed in the spectrum of RR~Tel (Thackeray
\cite{thackeray54}; Edlen \cite{edlen69}; Crawford et al. \cite{crawford99}); 
five of these lines ($\lambda$4868, $\lambda$4900, $\lambda$4903, $\lambda$4906 
and $\lambda$4918), from the same multiplet a$^{4}$G-a$^{4}$F, are within the 
spectral range of interest and were included in our model, improving the fit.

 We have identified the line [Ca~VII]$\lambda$4940 which had previously been 
detected in the spectrum of RR~Tel (Thackeray \cite{thackeray74}; McKenna et al.
\cite{mckenna97}). 

 Our fit requires a line at $\lambda$4932.5 ($\lambda$4932/$\lambda$5007=0.003);
Crawford (\cite{crawford99}) observed a line at $\lambda$4930.5 in the spectrum 
of RR~Tel and identified it with [O~III]$\lambda$4931.0; however the three 
[O~III]  lines at $\lambda$5007, $\lambda$4959 and $\lambda$4931 are from a 
triplet originating from the same upper level; the transition probabilities are 
such that $\lambda$4931/$\lambda$5007=9$\times$10$^{-5}$ while the value 
observed in RR~Tel is equal to 0.02, or 220 times larger than the theoretical 
value; we consider therefore that this line is unidentified.

\begin{table}[ht]
\caption{\label {narrow_lines} Narrow emission lines identified in the range
from 4780 to 5120 \AA\ . Col. 1: line identification, col. 2: laboratory 
wavelength, col. 3: relative observed intensity in HS~0328+05, col. 4: intensity
in RR~Tel (Crawford et al. 1999). }
\begin{center}
\begin{tabular}{lrrr}
\hline
 iden & $\lambda_{\rm lab}$ (\AA) & I$_{\rm HS}$ & I$_{\rm RR~Tel}$ \\
\hline
 $\rm [Fe~II]$ (20F) & 4814.5 &  0.69 &  0.367 \\
 H$\beta$            & 4861.3 & 32.10 & 30.550 \\
 $\rm [Fe~IV]$       & 4868.0 &  0.55 &  0.115 \\
 $\rm [Fe~II]$ (20F) & 4874.5 &  0.95 &  0.088 \\
 $\rm [Fe~III]$ (2F) & 4881.0 &  0.67 &  0.039 \\
 $\rm [Fe~II]$  (4F) & 4889.6 &  1.36 &  0.189 \\
 $\rm [Fe~VII]$ (2F) & 4893.4 &  0.21 &  1.250 \\
 ?                   & 4896.8 &  1.08 &        \\
 $\rm [Fe~IV]$       & 4900.0 &  0.27 &  0.056 \\
 $\rm [Fe~IV]$       & 4903.5 &  0.32 &  0.068 \\
 $\rm [Fe~II]$ (20F) & 4905.3 &  0.49 &  0.145 \\
 $\rm [Fe~IV]$       & 4906.7 &  0.75 &  0.185 \\
 $\rm [Fe~IV]$       & 4918.1 &  0.32 &  0.078 \\
 He~I (48)           & 4921.8 &  0.13 &  0.185 \\
 Fe~II (42)          & 4923.9 &  0.98 &  0.128 \\
 ?                   & 4932.5 &  0.88 &  0.873 \\
 $\rm [Ca~VII]$ (1F) & 4940.3 &  1.02 &  3.063 \\
 $\rm [Fe~VII]$ (2F) & 4942.5 &  0.43 &  2.472 \\
 $\rm [Fe~II]$ (20F) & 4947.4 &  0.67 &  0.039 \\
 $\rm [Fe~II]$ (20F) & 4950.7 &  0.32 &  0.056 \\
 $\rm [O~III]$ (1F)  & 4958.9 & 91.00 & 22.390 \\
 $\rm [Fe~VI]$ (2F)  & 4967.1 &  0.65 &  0.579 \\
 $\rm [Fe~VI]$ (2F)  & 4972.5 &  0.75 &  0.719 \\
 $\rm [Fe~II]$ (20F) & 4973.4 &  0.22 &        \\
 $\rm [Fe~VII]$ (2F) & 4988.6 &  0.54 &  1.854 \\
 $\rm [O~III]$ (1F)  & 5006.8 &273.00 & 39.360 \\
 He~I (4)            & 5015.7 &  0.75 &  0.328 \\
 Fe~II (42)          & 5018.4 &  0.70 &  0.180 \\
 $\rm [Fe~II]$ (20F) & 5020.2 &  0.34 &  0.086 \\
 $\rm [Fe~II]$ (20F) & 5043.5 &  0.38 &  0.031 \\
 ?                   & 5058.0 &  0.40 &        \\
 $\rm [Fe~II]$ (19F) & 5111.6 &  0.30 &  0.114 \\
 
\hline
\end{tabular}
\end{center}
\end{table}
\normalsize

 We have found another unidentified line at $\lambda$4896.8. There is also a 
weak line at $\sim\lambda$5058; Si~II $\lambda$5056 is not a good fit. 

 The narrow lines found in the range 4780 to 5120 \AA\ are listed in
Table \ref{narrow_lines}. They were all included in our fit assuming that they 
have the same velocity and width (4.9 \AA\ FWHM). \\

\begin{figure}[ht]
\resizebox{8.8cm}{!}{\includegraphics{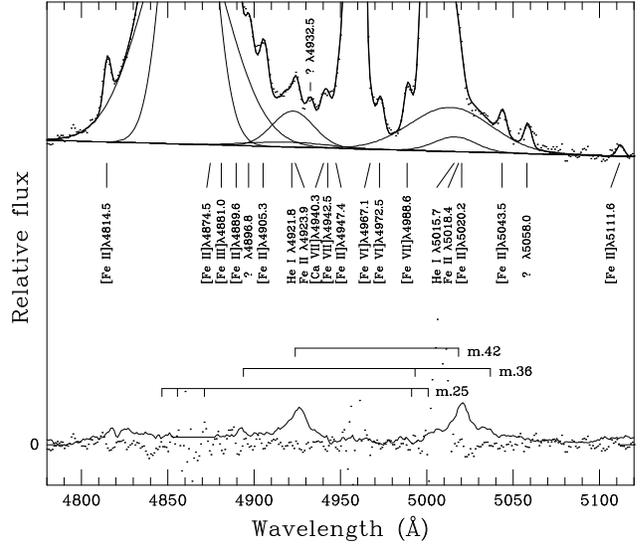}}
\caption{\label{hs_hb}
Fit of the spectrum of HS~0328+05 in the range 4780 to 5120 \AA. 
The figure shows the data (after subtraction of the Fe~II emission), the fit
and the broad individual components (H$\beta$ and He~I 
$\lambda\lambda$4922,5016). A number of narrow emission lines have been 
identified, many of them being due to iron. The ``red shelf" is well fitted by a
combination of narrow lines and  broad He~I $\lambda$4922 and $\lambda$5016 
components. The bottom solid line shows the Fe II spectrum subtracted from the 
original spectrum while the dotted line shows the residuals.}

\end{figure}

 In the range 4915 to 4955 \AA\ the emission excess is well fitted by a 
number of 
narrow emission lines ([Fe~IV]$\lambda$4918, He~I $\lambda$4922, Fe~II (42) 
$\lambda$4924, [Ca~VII] $\lambda$4940, [Fe~VII]$\lambda$4942, [Fe~II] (20F)
$\lambda\lambda$4947,4951 and an unidentified line at $\lambda$4932.5 and by the
broad components of He~I $\lambda$4922 (see Fig.~\ref{hs_hb}).
 We also find no need for a broad redshifted component to the [O~III] lines.

\section{Results and discussion}

\subsection{He I $\lambda$4922 and $\lambda$5016}

 Broad He~I emission lines are observed in AGNs: $\lambda$3188 (Baldwin et al. 
\cite{baldwin80}), $\lambda$4471 (van Groningen \& de Bruyn \cite{groningen};
Feldman \& MacAlpine \cite{feldman78}), $\lambda$5876 and $\lambda$7065 
(Feldman \& MacAlpine \cite{feldman78}; Morris \& Ward \cite{morris}) and 
$\lambda$10830 (LeVan et al. \cite{levan}). The He~I lines at $\lambda$4471, 
$\lambda$4922, $\lambda$5016, $\lambda$5876 and $\lambda$6678 have been detected
in the NLS1 galaxy Mark~110 (Kollatschny et al. \cite{kollatschny01}).

 The broad He~I $\lambda$4471 line is usually weak in Seyfert~1s 
($\lambda$4471/$\lambda$5876$\sim$0.1) implying n$_{\rm e}\sim$10$^{9}$
cm$^{-3}$ (Feldman \& MacAlpine \cite{feldman78}; Almog \& Netzer 
\cite{almog}). \\

\begin{figure}[ht]
\resizebox{8.8cm}{!}{\includegraphics{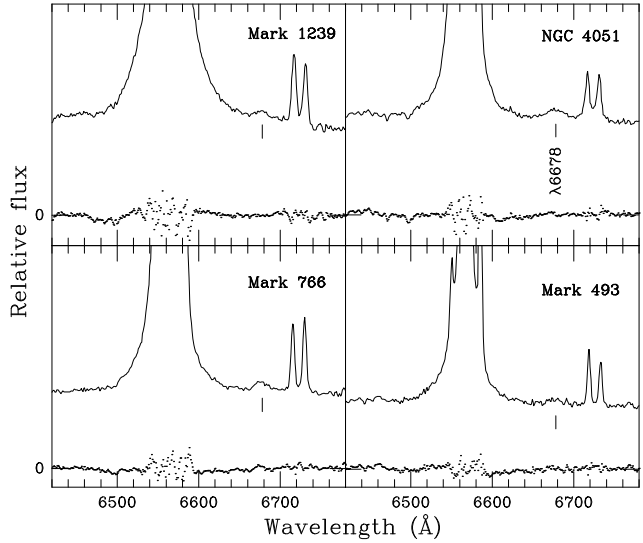}}
\caption{\label{he6678}
The He I $\lambda$6678 line observed in four NLS1s galaxies
}
\end{figure}

 Van Groningen \& de Bruyn (\cite{groningen}) concluded that the $\lambda$5016 
line emission could contribute only a small fraction of the ``red shelf"
observed in Seyfert 1s; this statement was based on the assumption that 
$\lambda$5016 should not be stronger than $\lambda$6678 and that the strength of
this last line is typically 0.6-0.8\% of that of H$\alpha$. Erkens et al. 
(\cite{erkens97}) have however observed broad $\lambda$6678 in a number of 
Seyfert~1s with an intensity relative to H$\alpha$ in the range 1.2-7.2\%.
 We have detected this line in several NLS1s with an intensity relative to 
H$\alpha$ in the range 0.8-2.6\% (Mark~1239: 0.8\%; NGC~4051: 2.6\%; Mark~766: 
1.7\%; Mark~493: 1.2\%) (see Fig.~\ref{he6678}). 

 In the case of Mark~110, Kollatschny et al. (\cite{kollatschny01}) convincingly
showed that the red wing of [O~III]$\lambda$5007 is due to a broad He~I 
$\lambda$5016 line. \\

 In the spectrum of RXS~J01177+3637 and HS~0328+05, we have found He~I lines 
emitted by the two broad Balmer line emission clouds; their relative intensities
are listed in Tables \ref{HeI_lines} and \ref{HS_HeI_lines}.

 In the case of HS~0328+05, we can set an upper limit to the intensity of a 
broad $\lambda$4471 component corresponding to $\lambda$4471/$\lambda$5876=0.15.

\begin{table}[ht]
\caption{\label {HeI_lines} Intensity relative to H$\alpha$ of the He~I lines 
observed in the spectrum of RXS~J01177+3637}
\begin{center}
\begin{tabular}{lrr}
\hline
      & \multicolumn{2}{c}{FWHM (km s$^{-1}$)} \\
      & 2150    & 6100 \\
\hline
  $\lambda$ (\AA) & I & I \\
\hline
 4713 &       &  6\% \\
 4922 &       & 10\% \\
 5016 & 1.8\% &  7\% \\
 5876 & 4.1\% & 12\% \\
 6678 & 2.0\% &  9\% \\
 7065 & 2.6\% &      \\

\hline
\end{tabular}
\end{center}
\end{table}
\normalsize

\begin{table}[ht]
\caption{\label {HS_HeI_lines} Intensity relative to H$\alpha$ of the He~I lines
observed in the spectrum of HS~0328+05}
\begin{center}
\begin{tabular}{lrr}
\hline
      & \multicolumn{2}{c}{FWHM (km s$^{-1}$)} \\
      & 1\,730    & 3\,400 \\
\hline
  $\lambda$ (\AA) & I & I \\
\hline
 4922 & 2.7\% &  1\% \\
 5016 & 1.1\% &  6\% \\
 5876 & 2.1\% &  5\% \\
 6678 & 0.9\% &      \\
 7065 & 2.4\% &      \\

\hline
\end{tabular}
\end{center}
\end{table}
\normalsize

 At very high densities (n$_{\rm e}>$10$^{13}$ cm$^{-3}$), the He~I lines may 
become very strong relative to the Balmer lines (Stockman et al. 
\cite{stockman77}); the Fe~II emission region has a density which can almost 
reach this value (10$^{10}<$n$_{\rm e}<$10$^{11}$-10$^{12}$ cm$^{-3}$) 
(Collin-Souffrin et al. \cite{collin80}, \cite{collin88}; Kwan et al. 
\cite{kwan95}). 

\subsection{H$\beta$ wing}

 The wing to the red of H$\beta$ has often been assumed to be H$\beta$ emission coming
from the same region as the [O~III] wings with a low $\lambda$5007/H$\beta$ 
ratio (close to unity) implying a relatively high density (10$^{7}$-10$^{8}$ 
cm$^{-3}$) (Meyers \& Peterson \cite{meyers}; van Groningen \& de Bruyn 
\cite{groningen}). Van Groningen \& de Bruyn (\cite{groningen}) found no 
evidence for major differences between the H$\beta$ and H$\gamma$ profiles in a
number of Seyfert 1s; they also found that the wings detected on the long 
wavelength side of the [O~III] lines have a clear counterpart in the Balmer 
lines.

 However, although the broad H$\beta$ line in Akn~120 has a red wing, the other 
Balmer lines do not show such a feature which, consequently, must be due to 
elements other than hydrogen (Foltz et al. \cite{foltz}; Doroshenko et al. 
\cite{doroshenko}). 

 In the spectra of RXS~J01177+3637 and HS~0328+05, neither H$\beta$ nor the 
[O~III] lines have a broad redshifted component.

\section{Conclusion}

 The detailed study of the spectrum of two Seyfert 1s showing a ``red 
shelf" shows that it is mainly due, in addition to the Fe~II multiplet 42, to 
the presence of relatively strong He~I 
$\lambda$4922 and $\lambda$5016 broad lines. In HS~0328+05, there is also 
a non negligible contribution to the H$\beta$ red wing of a number of narrow 
emission lines. 

 There is no evidence for the presence of a broad redshifted component in
H$\beta$ or [O~III] in any of these two objects.

\end{document}